# Calculation of Bit Error Ratio for

# Optically Pre-Amplified DPSK Receivers Using Optical Mach-Zehnder

# Interferometer Demodulation and Balanced Detection

*Xiupu Zhang, Zhenqian Qu, and Lei Wang,*

*Department of Electrical and Computer Engineering, Concordia University,*

*Montreal, Quebec, H3G 1M8, CANADA*

*Tel: 514 848 2424 -4107, Fax: 514 848 2802,*

*E-mail: xzhang@ece.concordia.ca*

**Abstract**: This paper presents an analysis of how to calculate bit error ratio (BER) with physical explanation for optically pre-amplified DPSK receivers using optical Mach-Zehnder interferometer (MZI) demodulation and balanced detection. It is shown that BER calculation method for this kind of receivers is different from the conventional calculation method used widely for IM/DD receivers. An analytical relationship in receiver sensitivity between DPSK receivers using MZI demodulation with balanced detection and IM/DD receivers (or DPSK receivers using MZI demodulation and single-port detection) is given based on the Gaussian noise approximation. Our calculation method correctly predicts the 3-dB improvement of receiver sensitivity by using balanced detection over single-port detection or IM/DD receivers. Furthermore, quantum-limited DPSK receivers with MZI demodulation are also analyzed.

## I. Introduction

Differential phase shifted keying (DPSK) is one of enabling techniques for the reduction of fiber Kerr nonlinearity in dense wavelength division multiplexing (DWDM) fiber transmissions [1]-[3]. Moreover, DPSK combined with optical Mach-Zehnder interferometer (MZI) demodulation and balanced detection (referred to DPSK/MZI receivers with balanced detection thereafter) provides a full optical demodulation and 3-dB improvement in receiver sensitivity over single-port detection or intensity modulation/direct detection (IM/DD) receivers (DPSK/MZI receivers with single-port detection is equivalent to IM/DD receivers if optically pre-amplified) [1-4]. In optically pre-amplified IM/DD receivers, noise statistic with optically amplified spontaneous emission (ASE) noise is characterized by the Chi-square distribution [5]. The Chi-square distribution is well approximated by the Gaussian distribution, which has been widely used for IM/DD receivers [5-8]. This is because ASE-ASE beat noise is only over-estimated by the Gaussian noise approximation and bit error ratio (BER) is thus slightly over-estimated by using Gaussian noise approximation. Since optical DPSK signal is converted into optical intensity modulated by MZI demodulator, the physical process of the signal and ASE noise in DPSK/MZI receivers is more close to IM/DD receivers than the conventional DPSK receivers which have electrical demodulation (consisting of an electrical time-delay and a mixer). Calculation of BER for optically pre-amplified IM/DD receivers has been well established and understood [5-8]. However, calculation of BER for optically pre-amplified DPSK/MZI receivers with balanced detection has not been fully understood so far. In such a receiver, both bits "1" and "0" have non-zero decision currents, rather than that only bit "1" has non-zero current in IM/DD receivers. This suggests that both bits "1" and "0" are detectable in DPSK/MZI receivers with balanced detection; and only bit "1" is detectable in IM/DD receivers in principle. Suppose that a bit "1" and a bit "0" are transmitted, and the bit "1" becomes zero-current due to some reasons in DPSK/MZI receivers with balanced detection and IM/DD receivers. Thus, an error occurs in IM/DD receivers since bit "0" always has zero-current and bits "1" and "0" are not distinguishable. In contrast, bit "0" has non-zero current in DPSK/MZI receivers with balanced detection, and bits "1" and "0" are still possible to be distinguished. This is the physical origin of why DPSK/MZI receivers with balanced detection provide 3-dB advantage over single-port detection or IM/DD receivers. Therefore, the same way as for IM/DD receivers in theoretical calculation of BER for DPSK/MZI receivers with balanced detection may not be correct [3], [8]. Recently, we used the exact probability density function (pdf) of noise statistics in optically pre-amplified DPSK/MZI receivers to calculate the cumulative error probability (CEP) by $CEP = \dfrac{1}{2}\left[\displaystyle\int_{-\infty}^{I_{th}} f_1\left(x\right)dx + \int_{I_{th}}^{\infty} f_0\left(x\right)dx\right]$, where $f_1\left(x\right)$ and $f_0\left(x\right)$ are the exact pdfs of noise statistics in bits "1" and "0", and $I_{th}$ is the optimal decision threshold current. It was verified that the ~3-dB improvement obtained previously in [3] [8] by using balanced detection is due to ASE-ASE beat noise in the two ports [9],[10].

On the other hand, noise statistics become the Gaussian if ASE-ASE beat noise is ignored, and ASE-ASE beat noise is only over-estimated by the Gaussian noise approximation in DPSK/MZI receivers with balanced detection as in IM/DD receivers [9]. As mentioned above, ~3-dB improvement by DPSK/MZI receivers with balanced detection over single-port detection or IM/DD receivers as shown in [3],[8] is due to two-port ASE-ASE beat noise, which induces the pdf shape deviated from the Gaussian distribution as shown in [9]. When signal-ASE beat noise is completely dominant (i.e. noise statistic becomes Gaussian), the ~3-dB improvement predicted in [3],[8] disappears if the above CEP is considered BER. Thus, DPSK/MZI receivers with balanced detection do not have 3-dB advantage and the performance of DPSK/MZI receivers with balanced detection or single-port detection and IM/DD receivers is identical. This does not converge with the measurements in which 3-dB improvement is always obtained no matter what signal power is used [1],[2],[4]. Moreover, the measured 3-dB improvement by balanced detection is well explained by signal constellation comparison [3], in which the signal energy used for error detection in DPSK/MZI receivers with balanced detection is double of that in single-port detection. Moreover, the signal constellation can be used for the two receiver comparison only when the two receivers have the same noise statistics. In other words, it is impossible to express the ASE-ASE beat noise induced pdf difference by using signal energy as shown in [3]. Consequently, the 3-dB improvement predicted in [3],[8], based on the conventional calculation method of $BER = \dfrac{1}{2}\left[\displaystyle\int_{-\infty}^{I_{th}} f_1\left(x\right)dx + \int_{I_{th}}^{\infty} f_0\left(x\right)dx\right]$ , is not the experimentally measured 3-dB improvement in [1-4]. Mathematically, the 3-dB improvement of receiver sensitivity is scaled to 3-dB Q factor *only if* the signal-ASE beat noise is dominating. If the predicted 3-dB in [3],[8] which is induced by ASE-ASE beat noise is the measured 3-dB, the above scale does not hold. But this scale always holds as shown in [1].

Recently, we proposed a calculation BER method and obtained 3-dB improvement for optically pre-amplified DPSK/MZI receivers with balanced detection [10] based on the Gaussian noise statistics. In this paper, we will show that how to explain and understand the calculation method of BER for DPSK/MZI receivers with balanced detection, and compare it with that for IM/DD receivers. Based on the Gaussian noise, the relationship of equivalent or effective Q-factor, for DPSK/MZI receivers with balanced detection, with Q factor for IM/DD receivers is given analytically.

## II. BER calculation method

The optically pre-amplified DPSK/MZI receiver with balanced detection is shown schematically in Fig. 1. The incoming optical signal from an optical pre-amplifier is expressed by an electric field $E_{in}(t)$. The optical filter is assumed an ideal filter and only used for filtering ASE noise, and the output from the optical filter is expressed in electric field $E_1(t)$. The optical MZI, which consists of two ideal 3-dB optical couplers and a piece of fiber used for time delay, converts the phase modulated signal into an amplitude modulated. The two outputs from the MZI are represented by electric fields $E_-(t)$ and $E_+(t)$, from the destructive and constructive ports, respectively. For an ideal MZI, the relationships between two outputs and input are given by $E_-(t) = \frac{1}{2}[E_1(t-T_b) - E_1(t)]$ and

$E_+(t) = \frac{-j}{2}[E_1(t-T_b) + E_1(t)]$, where $T_b$ is a bit period of the signal. The photodiode is modeled by a square-law detector with a responsivity of R and R=1 is assumed in this paper. The output current from photodiodes will pass a low-pass electrical filter (LPF) with the impulse response of $h_e(t)$ which is also assumed ideal and only used for noise filtering and no signal distortion induced.

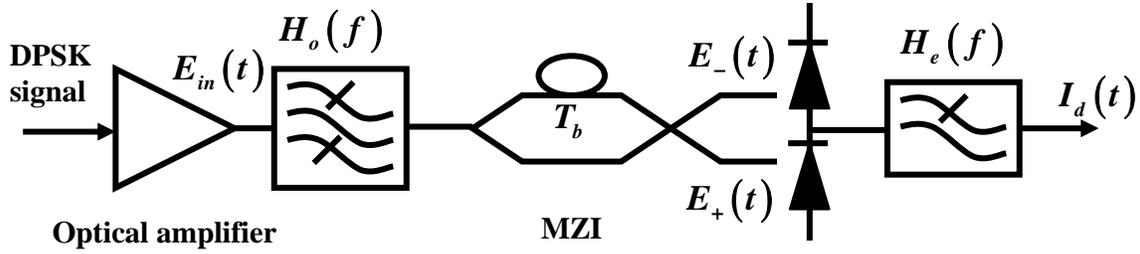

Fig. 1. Schematic blocks of DPSK/MZI receiver with balanced detection.

The electric field $E_{in}(t)$ of the optical signal output from the optical pre-amplifier can be expressed by

$$E_{in}(t) = E_0(t)e^{j\omega_0 t + j\theta(t)} + n(t)e^{j\omega_0 t} \qquad (1),$$

where

$\omega_0$ - Optical carrier frequency,

$E_0(t)$ - Amplitude modulation and assumed real without loss of generality;

$n(t)$ - Equivalent low-pass band ASE noise from the optical amplifier;

$\theta(t)$ - DPSK phase modulation of the signal.

The DPSK phase modulation, $\theta(t)$, can be written as $\theta(t) = \dfrac{\pi}{2}\sum_{k=-\infty}^{\infty} a_k g(t - kT_b) + \dfrac{\pi}{2}$ ,

where $g(t)$ is the modulation pulse shape and varies from 0 to 1 with time and $a_k$ is transmitted data, either "1" or "-1". For the *ideal* balanced detection, when bit "1" is received at the constructive port, only ASE noise shall present at the destructive port. When bit "0" is received at the destructive port, only ASE noise shall present at the constructive port. Therefore, the output currents $I_d(t)$ are given by [9]

$$I_1(t)/R = P_s + E_{s+}n_+^*(t) + E_{s+}^* n_+(t) + |n_+|^2 - |n_-|^2 \qquad (2a),$$

for bit "1", and

$$I_0(t)/R = -P_s - E_{s-}n_-^*(t) - E_{s-}^* n_-(t) - |n_-|^2 + |n_+|^2 \qquad (2b),$$

for bit "0". $E_{s+}$ ($n_+(t)$) and $E_{s-}$ ($n_-(t)$) denote output electric fields of signal (output ASE noise) at the constructive and destructive ports, respectively. $P_s$ is the average optical signal power. In (2) the second and third terms represent the signal-ASE beat noise and the last two terms represent the ASE-ASE beat noise from the two ports. Equation (2) can be simplified into

$$I_1(t) = I_s + n_1(t) \qquad (3a),$$

and $\qquad I_0(t) = -I_s - n_0(t) \qquad (3b),$

where $n_1(t)$ ($n_0(t)$) denotes the signal-ASE beat noise from the constructive port (destructive port), and ASE-ASE beat noise from both two ports; and $I_s = RP_s$ is the signal current at the decision time, which is corresponding to the average optical signal power. Equation (3) only holds for the *ideal* balanced detection.

Suppose that a bit "1" and a bit "0" are transmitted. Figure 2(a) shows the currents of bits "1" and "0" for a *noise free* DPSK/MZI receiver with balanced detection. For such a case, the decision threshold is set zero and no errors occur. Due to some reasons, the current of bit "0" is assumed to become positive but less than the current of bit "1" i.e. $I_0(t) < I_1(t)$, as shown in Fig. 2(b). By adjusting the decision threshold the bits "1" and "0" are still correctly detected (similar to IM/DD receivers). However, for the case as shown in Fig. 2(c) i.e. $I_0(t) > I_1(t)$, bit errors certainly occur, and the errors are totally induced by bit "0". Similarly, errors are induced by bit "1" only if $I_1(t) < I_0(t)$. Thus, for the noisy DPSK/MZI receivers with the *ideal* balanced detection BER can be calculated by $BER_1 = \mathrm{Prob}[I_1 < I_0]$ from bit "1", and $BER_0 = \mathrm{Prob}[I_0 > I_1]$ from bit "0", i.e. [10]

$$BER_0 = \int_{-\infty}^{\infty} f_1(x) \left[ \int_{x}^{\infty} f_0(y) dy \right] dx = \frac{1}{2\pi\sigma^2} \int_{-\infty}^{\infty} \exp\left[ -\frac{(x-I_s)^2}{2\sigma^2} \right] dx \int_{x}^{\infty} \exp\left[ -\frac{(y+I_s)^2}{2\sigma^2} \right] dy$$

$$= \frac{1}{\sqrt{2}\sigma\sqrt{2\pi}} \int_{0}^{\infty} \exp\left[ -\frac{(x+2I_s)^2}{2(2\sigma^2)} \right] dx$$

$$(4a),$$

and

$$BER_1 = \int_{-\infty}^{\infty} f_0(x) \left[ \int_x^{\infty} f_1(y) dy \right] dx = \frac{1}{2\pi\sigma^2} \int_{-\infty}^{\infty} \exp\left[ -\frac{(x+I_s)^2}{2\sigma^2} \right] dx \int_x^{\infty} \exp\left[ -\frac{(y-I_s)^2}{2\sigma^2} \right] dy$$

$$= \frac{1}{\sqrt{2}\sigma\sqrt{2\pi}} \int_{-\infty}^{0} \exp\left[ -\frac{(x-2I_s)^2}{2(2\sigma^2)} \right] dx$$

(4b),

where $\sigma^2 = \sigma_1^2 = \sigma_0^2$ was used and $\sigma_1^2$ ($\sigma_0^2$) is the variance of the noise $n_1(t)$ ($n_0(t)$). In (4) the Gaussian noise of $n_1(t)$ and $n_0(t)$ were assumed in the last two steps. It is clearly observed for the ideal balanced detection that the decision threshold in current is always zero, the equivalent means of the bit "0" and "1" currents are $-2I_s$ and $2I_s$, respectively, and the equivalent variances of the bit "0" and "1" currents are the same and equal to $2\sigma^2$.

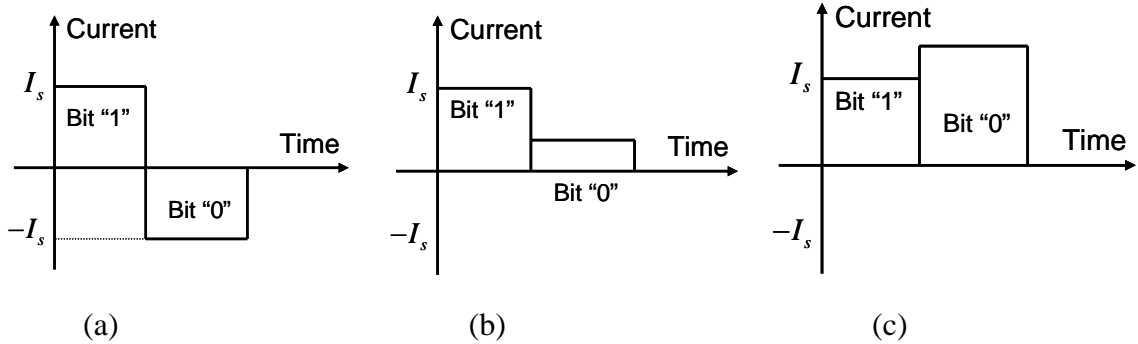

(a)　　　　　　　　(b)　　　　　　　　(c)

Fig. 2. Signal currents of bits "1" and "0" in noise free DPSK/MZI receivers with balanced detection for three cases: (a) an ideal case, (b) a non-deal case with the current of bit "0" $I_0(t) > 0$ and $I_0(t) < I_1(t)$, (c) a non-deal case with the current of bit "0" $I_0(t) > 0$ and $I_0(t) > I_1(t)$.

Alternatively, expressions (4) can be obtained as follows. Because the currents for bits "1" and "0" have the opposite sign in DPSK/MZI receivers with balanced detection, error detection for bit "1" is made by the current $I_1(t)$ with the reference to bit "0" current $I_0(t)$, and error detection for bit "0" is made by the current $I_0(t)$ with the reference to bit "1" current $I_1(t)$. In other words, bit "1" is detected based on the current of bit "1" with the reference of bit "0" current, and vice versa. Therefore, the equivalent decision variables are $y_1 = I_1(t) - I_0(t)$ for bit "1" and $y_0 = I_0(t) - I_1(t)$ for bit "0". Thus, the errors occur from bits "1" and "0" if the equivalent decision variables satisfy $y_1 = I_1(t) - I_0(t) < I_{th} = 0$ and $y_0 = I_0(t) - I_1(t) > I_{th} = 0$, respectively ($I_{th} = 0$ only for the ideal balanced detection). It is easily obtained that the means of the equivalent decision variables $y_1$ and $y_0$ in current are $2I_s$ and $-2I_s$, respectively, and the variances of $y_1$ and $y_0$ are the same, i.e. $2\sigma^2$. Based on these facts and Gaussian noise statistics, we can easily obtain the expressions (4).

For IM/DD receivers, the decision currents are $I_1(t) = 2I_s + n_1(t)$ for bit "1" and $I_0(t) = n_0(t)$ for bit "0". We have already assumed that the peak power of bit "1" is twice of the average optical power and thus optical/electrical signal to noise ratio is the same in the two receivers. $n_1(t)$ is noise with the variance $\sigma_1^2$ in bit "1", consisting of signal-ASE beat noise and ASE-ASE beat noise. $n_0(t)$ is noise with the variance $\sigma_0^2$ in bit "0", consisting of ASE-ASE beat noise only. Because the currents of bits "1" and "0" both are positive with the minimum current of zero, the best reference to make decisions for both bits "1" and "0" is zero. Then, errors occur when the decision variables $y_1 = I_1(t) - 0 < I_{th}$ for bit "1", and $y_0 = I_0(t) - 0 > I_{th}$ for bit "0". Based on these facts and Gaussian noise statistics, the well-known expressions of BER are given by [5-8],

$$BER_0 = \frac{1}{\sigma_0 \sqrt{2\pi}} \int_{I_{th}}^{\infty} \exp\left[-\frac{x^2}{2\sigma_0^2}\right] dx, \qquad (5a),$$

and

$$BER_1 = \frac{1}{\sigma_1 \sqrt{2\pi}} \int_{-\infty}^{I_{th}} \exp\left[-\frac{(x - 2I_s)^2}{2\sigma_1^2}\right] dx \qquad (5b).$$

By comparing BER expressions of (4) and (5), it is seen that the expressions of BER for both bits "1" and "0" are very similar, and the only differences are the equivalent means and variances in DPSK/MZI receivers with balanced detection and IM/DD receivers, besides that the decision threshold in DPSK/MZI receivers with *ideal* balanced detection is always zero, independent of the means and variances.

## III. Relationship of Q factors

Since the ASE-ASE beat noise is only over-estimated by the Gaussian noise approximation in both IM/DD and DPSK/MZI receivers with balanced detection, particularly for the last ones [9], BER calculated by the Gaussian noise is over-estimated. However, the Gaussian noise approximation still provides us with a fair estimation of BER since signal-ASE beat noise is usually dominating and has the Gaussian noise. Therefore, we analyze Q-factor for IM/DD receivers and DPSK/MZI receivers with balanced detection based on the simple Gaussian noise. For IM/DD receivers, it is well known and given by $Q_{IM/DD} = \dfrac{2I_s - 0}{\sigma_1 + \sigma_0}$ [5]-[8],[10] by using approximated optimal threshold†, where $\sigma_1^2 = \sigma_{SA,IM/DD}^2 + \sigma_{AA,IM/DD}^2 = 8I_s(RN_{ASE})B_e + 2R^2N_{ASE}^2(2B_o - B_e)B_e$ is the variance of the noise $n_1(t)$ for bit "1", and $\sigma_0^2 = \sigma_{AA,IM/DD}^2 = 2R^2N_{ASE}^2(2B_o - B_e)B_e$ is the variance of the noise $n_0(t)$ for bit "0", $\sigma_{SA,IM/DD}^2$ - the variance of the signal-ASE beat noise, $\sigma_{AA,IM/DD}^2$ -the variance of the ASE-ASE beat noise, $N_{ASE}$ -ASE noise spectral density in one polarization state, $B_o$ -optical noise bandwidth, and $B_e$ -electrical noise bandwidth

(† *the optimal decision threshold should be determined by* $BER_1 = BER_0$ *rather than by the probability density functions* $f_1(x) = f_0(x)$ *in [5-8].)*

For DPSK/MZI receivers with the *ideal* balanced detection, the equivalent or effective Q factor is the same for bits "1" and "0", and given by $Q_{DPSK} = \dfrac{2I_s - I_{th}}{\sqrt{2}\sigma} = \dfrac{2I_s - 0}{\sqrt{2}\sigma}$ for bit "1", and $Q_{DPSK} = \dfrac{I_{th} - (-2I_s)}{\sqrt{2}\sigma} = \dfrac{2I_s}{\sqrt{2}\sigma}$ for bit "0", where $\sigma^2 = \sigma_{SA,DPSK}^2 + \sigma_{AA,DPSK}^2 = 2I_s(RN_{ASE})B_e + R^2N_{ASE}^2(B_o - B_e)B_e$ is the variance of $n(t)$, $\sigma_{SA,DPSK}^2$ -the variance of the signal-ASE beat noise and $\sigma_{AA,DPSK}^2$ -the variance of the ASE-ASE beat noise (*see Appendix A*). After simple algebras, we can obtain the relationship of Q factors between IM/DD receivers and DPSK/MZI receivers with balanced detection,

$$Q_{DPSK} = \sqrt{2} Q_{IM/DD} \frac{\sqrt{1 + \frac{\sigma_{AA,IM/DD}^2}{4\sigma_{SA,DPSK}^2}} + \frac{\sigma_{AA,IM/DD}}{2\sigma_{SA,DPSK}}}{\sqrt{1 + \frac{\sigma_{AA,DPSK}^2}{\sigma_{SA,DPSK}^2}}} \qquad (6).$$

It is shown by (6) that the exact 3-dB improvement of Q factor by DPSK/MZI receivers with *ideal* balanced detection over IM/DD receivers is achieved if only including the signal-ASE beat noise in the two receivers. For the case of ASE-ASE beat noise included, an additional improvement of ~1 dB due to ASE-ASE beat noise is obtained by using the Gaussian noise approximation for the typical receiver bandwidths, instead of ~3 dB by using the exact noise statistics given in [9]. Therefore, DPSK/MZI receivers with balanced detection ultimately outperform IM/DD receivers or DPSK/MZI receivers with single-port detection by exact 3 dB. In [3] the 3-dB improvement is interpreted by signal constellation. Our predicted 3-dB improvement by using balanced detection converges with the interpretation of the 3-dB improvement by signal constellation.

One serious argument is that bits "1" and "0" never appear at the same time physically, therefore BER calculation cannot be based on the conditions of $y_1 = I_1(t) - I_0(t) < 0$ and $y_0 = I_0(t) - 0 > I_{th}$ [11] because either bit "1" or "0" only appears at the decision instant. In other words, when bit "1" is being detected bit "0" is not known and vice versa. Before we answer this argument we first review BER calculation for IM/DD receivers. It is well known that BER calculation for IM/DD receivers is based on the optimal decision threshold [5]-[8]. We also know that the optimal decision threshold is calculated by using the decision currents and variances of both bits "1" and "0" [5]-[8]. If bit "1" is being detected and bit "0" is not known, BER for IM/DD receivers cannot be calculated based on the optimal decision threshold because bit "0" is not known and it is impossible to know the optimal decision threshold. Moreover, Q-factor for IM/DD receivers is also calculated based on the optimal decision threshold. This suggests that BER and Q-factor calculations for IM/DD receivers are incorrect. However, calculated BER and Q-factor based on the optimal decision threshold for IM/DD receivers have been verified experimentally and used for tens of decades. Consequently, the understanding of which when bit "1" is being detected and bit "0" is not known and vice versa in BER calculation is incorrect.

### IV. Conclusions

We have, for the first time, presented an analysis of how to calculate BER and provided physical explanation of BER calculation for DPSK/MZI receivers with balanced detection. The simple relationship of Q factors for DPSK/MZI receivers with balanced detection and IM/DD receivers is given based on the Gaussian noise approximation. The predicted improvement of 3-4 dB based on the Gaussian noise agrees well with the measured [1]-[4]. Moreover, our predicted 3-dB improvement has no conflict with the signal constellation.

### Appendix A: Variance of ASE-ASE beat noise

In this appendix, we analyze ASE-ASE beat noise for DPSK/MZI receivers with balanced detection. Supposed that the optical filter before the MZI is an ideal filter, i.e. $H_o(f) = 1$ for $-B_o/2 < f < B_o/2$ and $H_o(f) = 0$ elsewhere, $B_o$ - the noise bandwidth of the optical filter. The frequency responses of the ideal MZI are given by

$$H_{MZI}(f) = \begin{cases} \cos(\pi f T_b)\exp(j\pi f T_b) & \text{cosntructive port} \\ \sin(\pi f T_b)\exp(j\pi f T_b) & \text{destructive port} \end{cases} \tag{A1}$$

Thus, the equivalent noise bandwidths are given by

$$B_{o,c} = \int_{-\infty}^{\infty} \cos^2(\pi f T_b)\left|H_o(f)\right|^2 df = \int_{-B_o/2}^{B_o/2} \cos^2(\pi f T_b) df = \frac{B_o}{2} + \frac{R_b}{2\pi}\sin(\pi B_o T_b) \tag{A2a}$$

for the constructive port; and

$$B_{o,d} = \int_{-\infty}^{\infty} \sin^2(\pi f T_b)\left|H_o(f)\right|^2 df = \int_{-B_o/2}^{B_o/2} \sin^2(\pi f T_b) df = \frac{B_o}{2} - \frac{R_b}{2\pi}\sin(\pi B_o T_b) \tag{A2b}$$

for the destructive port, where $R_b$ is the bit rate.

For DPSK/MZI receivers with balanced detection, the variance of ASE-ASE beat noise is given by

$$\sigma_{AA,DPSK}^2 = 2R^2 \left(\frac{N_{ASE}}{2}\right)^2 \int_{-\infty}^{\infty} df \left\{ \left| H_{o,c}(f) \right|^2 \otimes \left| H_{o,c}(-f) \right|^2 \right\} \left| H_e(f) \right|^2$$

$$+ 2R^2 \left(\frac{N_{ASE}}{2}\right)^2 \int_{-\infty}^{\infty} df \left\{ \left| H_{o,d}(f) \right|^2 \otimes \left| H_{o,d}(-f) \right|^2 \right\} \left| H_e(f) \right|^2 \quad \text{(A3)}$$

$$\approx R^2 \frac{N_{ASE}^2}{2} \left(2B_{o,c} - B_e\right) B_e + R^2 \frac{N_{ASE}^2}{2} \left(2B_{o,d} - B_e\right) B_e$$

where $H_{o,c}(f)$ and $H_{o,d}(f)$ are the frequency responses of the constructive and destructive ports with the noise bandwidths of $B_{o,c}$ given by (A2a) and $B_{o,d}$ given by and (A2b), respectively. By using (A2a) and (A2b) in (A3), we obtain

$$\sigma_{AA,DPSK}^2 = R^2 N_{ASE}^2 \left(B_o - B_e\right) B_e \quad \text{(A4)}$$

For DPSK/MZI receivers with single-port detection, the variance of ASE-ASE beat noise is given by,

$$\sigma_{AA}^2 = 2R^2 \left(\frac{N_{ASE}}{2}\right)^2 \int_{-\infty}^{\infty} df \left\{ \left| H_{o,e}(f) \right|^2 \otimes \left| H_{o,e}(-f) \right|^2 \right\} \left| H_e(f) \right|^2$$

$$\approx R^2 \frac{N_{ASE}^2}{2} \left(2B_{o,e} - B_e\right) B_e \quad \text{(A5)}$$

where $H_{o,e}(f)$ is the equivalent optical frequency response of constructive or destructive port, and $B_{o,e}$ is the equivalent noise bandwidth of constructive or destructive port, given by (A2).

## Appendix B: Quantum limited DPSK/MZI receivers

In this Appendix we present BER analysis for DPSK/MZI receivers *considering quantum noise (shot noise) only*. The quantum limited receiver sensitivity of DPSK receivers, given by $BER = \frac{1}{2} \exp\left(-\bar{N}_p\right)$ $\bar{N}_p$ - the photon number per bit, has been widely used for both DPSK/MZI receivers with single-port or balanced detections [3],[4]. The above quantum limited BER was obtained for DPSK with electrical demodulation (referred to the conventional DPSK receiver thereafter), which consists of an electrical time delay line and an electrical mixer, based on Rice (bit "1") and Rayleigh (bit "0") noise statistics [12],[13]. However, the optical MZI demodulator in DSPK/MZI receivers converts DPSK optical signal into intensity modulated before the injection to the optical photodiodes. The noise statistic of quantum noise in DPSK/MZI receivers does not have the Rice and Rayleigh probability distributions; and as a matter of fact the **Gaussian/Poisson** noise distribution should be used as in IM/DD receivers [6, pp.167]. Moreover, DPSK/MZI receivers with balanced detection could be different from DPSK/MZI receivers with single-port detection in quantum limited receiver sensitivity, because the signal energy used for error detection is different in the two detections. Consequently, it could be expected that the quantum limited BER for DPSK/MZI receivers with single-port or balanced detections may be different from that of the conventional DPSK receivers. In this Appendix, we present a quantum limited analysis for DPSK/MZI receivers with single-port or balanced detection.

### B.1 Definitions of quantum- and quasi-quantum noise

When quantum noise is only considered, a small number of photons and electron-hole pairs present (i.e., the number of photons and electrons are countable). The noise statistics for DPSK/MZI receivers should follow the Poisson distribution (a discrete probability distribution) as in IM/DD receivers [6, pp.167]. As the number of photons and electrons becomes large enough, the noise statistics can be characterized by the Gaussian distribution (a continuous probability distribution). In this Appendix, quasi-quantum limited (QQL) analysis is referred if the quantum noise is considered to be the Gaussian noise, to distinguish it from the quantum limited (QL) analysis in which the quantum noise is considered to be Poisson noise. For the conventional DPSK receivers, BER expression of $BER = \frac{1}{2}\exp\left(-\bar{N}_p\right)$ [12],[13] is corresponding to our defined quasi-quantum limited analysis because the continuous Rice and Rayleigh noise statistics are used.

### B.2 Quantum limited analysis

We first analyze the quantum limited (Poisson noise statistics) DPSK/MZI receivers. We first consider DPSK/MZI receivers with single-port detection. If $m_1 > 0$ electron-hole pairs with the Poisson probability of $P(m_1) = \exp\left[-\bar{N}_p\right] \bar{N}_p^{m_1} / m_1!$ are generated by photon number $\bar{N}_p$ ($\bar{N}_p$ - the photon number in bit "1", and corresponding to the average optical power of the DPSK signal), no errors from bit "1" occur. Since bit "0" has zero photons and noise free, bit "0" is not detectable and BER is totally determined by bit "1" (Note $m_0 \equiv 0$ the number of electron-hole pairs in bit "0") similar to IM/DD receivers [6, pp.167]. Therefore the quantum limited BER is given by setting $m_1 = 0$ in the above Poisson distribution similar to IM/DD receives [6, pp.167], i.e.

$$BER_{S-QL} = \frac{1}{2}\left[\exp\left(-\bar{N}_p\right) + 0\right] \tag{B1}.$$

The receiver sensitivity given by (B1) is 3-dB worse than that in IM/DD receivers [6, pp.167] ($BER_{IM/DD} = \frac{1}{2}\exp\left(-2\bar{N}_p\right)$, the peak power of bit "1" is assumed twice the average power). This can be explained that only the half signal energy is used for error detection in DPSK/MZI receivers with single-port detection rather than the full signal energy in IM/DD receivers.

For DPSK/MZI receivers with balanced detection, bits "1" and "0" contain the same number of photons. When bit "1" is transmitted, no errors occur if $m_1 > 0$ electron-hole pairs with probability of $P(m_1) = \exp\left[-\bar{N}_p\right] \bar{N}_p^{m_1} / m_1!$ are generated at the constructive port. Similarly, no errors occur from bit "0" if $m_0 > 0$ electron-hole pairs with probability of $P(m_0) = \exp\left[-\bar{N}_p\right] \bar{N}_p^{m_0} / m_0!$ are created at the destructive port. In addition, let us consider the special case of $m_1 > 0$ and $m_0 = 0$. This cases is exactly the same as for DPSK/MZI receivers with single-port detection, in which no errors occur if $m_1 > 0$ and $m_0 = 0$. Therefore, no errors occur for this case and vice versa. Consequently, no errors occur if $m_0 + m_1 > 0$. In other words, an error shall occur only if $m_0 + m_1 = 0$ with the probability of $P\left(m = m_0 + m_1\right) = \exp\left[-2\bar{N}_p\right]\left(2\bar{N}_p\right)^m \Big/ m!$. Thus, BER for DPSK/MZI receivers with balanced detection is given by setting $m = 0$,

$$BER_{B-QL} = \frac{1}{2}\exp\left[-2\bar{N}_p\right] \qquad (B2).$$

The factor 1/2 is due to two bits. By comparing (B1) and (B2), we can find that the 3-dB quantum limited receiver sensitivity is improved by DPSKMZI receivers with balanced detection over single-port detection. On the other hand, the same quantum limited BER for DPSK/MZI receivers with balanced detection as IM/DD receivers ($BER_{IM/DD} = \frac{1}{2}\exp\left(-2\bar{N}_p\right)$) is obtained. This is because the two receivers use the same signal energy for error detection and also have the same total noise variance. The expression (B2) is given for the first time. It is shown that the quantum limit BERs are different for DPSK/MZI receivers with single-port or balanced detection. Therefore, it is not correct to use the expression (B1) for DPSK/MZI receivers with balanced detection [3]. If non-ideal photodiodes are considered, $\eta\bar{N}_p$ should replace $\bar{N}_p$ in (B1) and (B2), $\eta$ - the quantum efficiency of the photodiodes.

Furthermore, it is observed that the expression (B1) for DPSK/MZI receivers with single-port detection is the same as the conventional DPSK receivers. However BER given by (B1) is obtained based on the discrete Poisson distribution, rather than the continuous Rice and Rayleigh distributions. Particularly, it is worth emphasizing that the quantum-limited BER expression of $BER = \frac{1}{2}\exp\left(-\eta\bar{N}_p\right)$, which has been widely used for DPSK receivers [6],[8],[13], is only correct for DPSK/MZI receivers with single-port detection and the conventional DPSK receivers.

### B.3 Quasi-quantum limited analysis

We now start the analysis for the quasi-quantum limited (Gaussian noise statistics) DPSK/MZI receivers. For DPSK/MZI receivers (either single-port or balanced detections), the electrical signal-to-noise ratio is the same as that in IM/DD receivers provided that the average optical power in DPSK/MZI and IM/DD receivers is the same.

First let's consider DPSK/MZI receivers with single-port detection. The decision current $I_1(t)$ for bit "1" is corresponding to the average optical power rather than the peak power in IM/DD receivers. The decision currents for DPSK/MZI receivers with the constructive-port detection are $I_1(t) = R\overline{P}_s + n_1(t)$ for bit "1" and $I_0(t) = 0$ for bit "0", where R is the responsivity of the photodiodes, $\overline{P}_s$ denotes the average optical power, and $n_1(t)$ is the quantum noise with the variance of $\sigma^2$. The quasi-quantum limited BER for DPSK/MZI receivers with single-port detection is similar to IM/DD receivers [6, pp.167],

$$BER_{S-QQL} = \frac{1}{2} erfc\left[\frac{I_s}{\sigma\sqrt{2}}\right] = \frac{1}{2} erfc\left(\sqrt{\frac{\eta\overline{N}_p}{2}}\right) \qquad (B3),$$

where $erfc()$ is the complementary error function. In (B3) $I_s = R\overline{P}_s$, and $\sigma^2 = 2eI_s B_e$, the shot noise for bits "1", $e$ - electron charge, $B_e$ - the electrical noise bandwidth were used. For $B_e$ equal to the half of the bit rate, we obtain $\frac{I_s^2}{\sigma^2} = \eta\overline{N}_p$, which was used in the last step of (B3). It is seen that DPSK/MZI receivers with single–port detection is 3-dB worse than IM/DD receivers in receiver sensitivity ($BER_{IM/DD} = \frac{1}{2} erfc\left[\sqrt{\frac{\eta 2\overline{N}_p}{2}}\right]$

[6, pp.167]), again the same conclusion as the quantum-limited analysis.

For DPSK/MZI receivers with balanced detection, the decision currents are $I_1(t) = R\overline{P}_s + n_1(t)$ for bit "1" and $I_0(t) = -R\overline{P}_s - n_0(t)$ for bit "0". The quantum noise $n_0(t)$ and $n_0(t)$ of bits "1" and "0" have the same variance with $\sigma^2 = 2eI_s B_e$. Similar to the above, an error occurs only if $I_1(t) - I_0(t) < 0$ from bit "1". Thus, BER can be obtained by,

$$BER_{B-QQL} = \Pr ob(I_1 < I_0) = \int_{-\infty}^{\infty} \frac{1}{\sqrt{2\pi}\sigma} \exp\left[-\frac{(x-I_s)}{2\sigma^2}\right] dx \int_{x}^{\infty} \frac{1}{\sqrt{2\pi}\sigma} \exp\left[-\frac{(y+I_s)}{2\sigma^2}\right] dy$$

$$= \frac{1}{2} erfc\left(\frac{I_s}{\sigma}\right) = \frac{1}{2} erfc\left(\sqrt{\eta\overline{N}_p}\right) \qquad (B4).$$

In (B4) the same conditions as in (B3) have been applied in the last step. By comparing (B3) and (B4), it is found that BER given by (B3) and (B4) differs from 3-dB in receiver sensitivity. In other words, 3-dB receiver sensitivity is improved by DPSK/MZI receivers with balanced detection over single-port detection in the quantum limit. On the other hand, DPSK/MZI receivers with balanced detection has the same quantum limit as IM/DD receivers, since the total signal energy and noise variance, used for error detection in DPSK/MZI receivers with balanced detection, is exactly the same as in IM/DD receivers. The BER expressions of (B3) and (B4) are different from the expression of $BER = \frac{1}{2} \exp\left(-\eta \bar{N}_p\right)$ obtained for the conventional DPSK receivers based on the continuous Rice and Rayleigh distributions [12],[13].

### B.4 Summary

In Appendix B we have presented an analysis of DPSK/MZI receivers with single-port or balanced detections, considering the quantum noise only. We have found that 3-dB quantum limited receiver sensitivity differs between DPSK/MZI receivers with balanced detection and single-port detection. Moreover, DPSK/MZI receivers with balanced detection has the same quantum limit as IM/DD receivers, since the total signal energy and noise variance for error detection in both receivers are the same. The quantum limited BER with $BER_s = \frac{1}{2} \exp\left(-\eta \bar{N}_p\right)$ for DPSK/MZI receivers with single-port detection and $BER_b = \frac{1}{2} \exp\left(-2\eta \bar{N}_p\right)$ for balanced detection are given for the first time, based on the Poisson statistic. BER expressions and receiver sensitivity improvement of quantum noise limited DPSK/MZI receiver and conventional DPSK receiver compared to IM/DD receiver are summarized in Table 1. It is worthy to be emphasized that if the conventional BER calculation method $BER = \frac{1}{2}\left[\int_{-\infty}^{I_{th}} f_1\left(x\right) dx + \int_{I_{th}}^{\infty} f_0\left(x\right) dx\right]$ is used the exact same BER will be obtained for DPSK/MZI receivers with both single-port and balanced detection, and both receivers have 3-dB receiver sensitivity worse than IM/DD receivers.

**Table 1** BER expressions and receiver sensitivity improvement of quantum noise limited DPSK/MZI receiver and conventional DPSK receiver compared to IM/DD receiver

| Receiver | Decision current | Quantum noise limited | Quasi-Quantum noise limited | Improvement [dB] |
|---|---|---|---|---|
| IM/DD [6][13] | 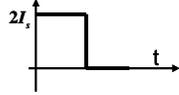 | $\frac{1}{2}\exp\left[-2\bar{N}_p\right]$ | $\frac{1}{2}erfc\left[\sqrt{\bar{N}_p}\right]$ | -- |
| Conventional DPSK [6][13] | 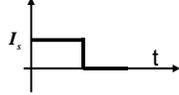 | - | $\frac{1}{2}\exp\left[-\bar{N}_p\right]$ | -3* |
| DPSK/MZI single-port | 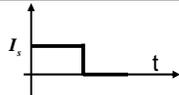 | $\frac{1}{2}\exp\left[-\bar{N}_p\right]$ | $\frac{1}{2}erfc\left[\sqrt{\frac{\bar{N}_p}{2}}\right]$ | -3 |
| DPSK/MZI balanced | 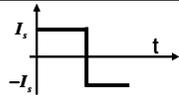 | $\frac{1}{2}\exp\left[-2\bar{N}_p\right]$ | $\frac{1}{2}erfc\left[\sqrt{\bar{N}_p}\right]$ | 0 |

The quasi-quantum noise limited conventional DPSK receiver is usually compared to the quantum noise limited IM/DD receiver [6][13].

Dear Dr. Zhang:

We have received the reviewers' comments on your manuscript JLT-08703-2005, "Calculation of Bit Error Ratio for Optically Pre-Amplified DPSK Receivers Using Optical Mach-Zehnder Interferometer Demodulation and Balanced Detection."   I regret to say that, based on the attached reviews, this paper does not meet our publication requirements.

Although we may have misjudged some aspect of this paper that you consider important, we do not encourage resubmission.   The extra time delays and demands on reviewers are counter to our objective to publish important new results rapidly.

Nevertheless, thank you for allowing us to examine the paper, and I hope you will consider submitting new manuscripts to Journal of Lightwave Technology in the future.

Sincerely,

Douglas M. Hargis, Publications Coordinator

IEEE/OSA Journal of Lightwave Technology

445 Hoes Lane

Piscataway, NJ 08854 USA

Reviewer 1 Comments:

    The presented paper describes the BER expression for optical amplified DPSK receivers. This subject was intensively studied in recent years, and then a number of trials have been carried out to address this issue, to date. The claim in the

submitted paper has been fundamentally well known in the several publications. Therefore, the reviewer does not find sufficient originality. The prior publications are listed as below;

Section II

The BER expression for DPSK and OOK is seen in for example; - Stein, Seymour & J.Jay Jones, �Modern Communication Principles,� McGraw Hill, 1967-.

Section III

A modified Q factor was already proposed in � C. Xu, X. Liu; X. Wei, �Differential phase-shift keying for high spectral efficiency optical transmissions,� J. Selected Topics in Quantum Electron, vol. 10, no. 2, pp. 281- 293, 2004-.

Unless the authors drastically change the main body clear to readers that the claim is sufficiently supported with sufficient originality, the reviewer has to say that the submitted paper cannot be accepted.

Reviewer 2 Comments:

The authors propose a paper dealing with the calculation of BER in optically pre-amplified DPSK receivers using optical Mach-Zehnder Interferometer demodulation and balanced detection.

As said in the abstract, the purpose of the study is the derivation of a modified Q-factor for DPSK systems, linked with the conventional and well-known Q-factor in IM/DD systems.

The authors derive a new form for Q-factor (amplitude Gaussian noises) in DPSK systems and confirm the 3 dB gain performed with the balanced detection compared to the single ended detection. I think that it is a good start but should be completed by:

1. Recently, some works [1] reports that conventional Q-factor is not appropriate to calculate the BER of DPSK systems, and typically the conventional method underestimates the performance (BER) in the linear regimes and overestimates the BER in the high non linear regimes (figure 10 in ref[1]). As a new Q-factor is proposed, could the author obtain such a curve (figure 10 in ref[1]) and compare their results on this point?

2. All the pdfs used for noises are assumed to be Gaussian. Does it is sufficient to obtain a good estimation of the performances in DPSK systems? If yes could you justify it?

3. The expression of theta(t) (page 5) given for the DPSK signal assumes that the DPSK signal has no phase noise at the receiver. As Kerr effect in optical fiber degrades DPSK signals along propagation and induce non linear phase noise, I think that theta(t) should contain a noise contribution. What is the justification of authors to this point?

4. The interest of Appendix B is not clearly showed: as the authors don�t mention it in the main text, could you precise the main interest of this part for your work (could you include it in the main text?). As the results of BER are expressed (table 1) with the erfc function (limit of the BER = 0 for high values of Np), how the non-linear regime with these formulas can be considered (compared with figure 10 in ref[1])?

According to these remarks, I need some precisions to give a valuable judgement of this paper.

[1] IEEE Journal of Selected Topics in Quantum Electronics, Vol. 10, No. 2, March/April 2004, Xu and al., �DSPK for High Spectral Efficiency Optical Transmissions�.

##editor_to_author##

My decision is based on my own study of the manuscript and the reviews.   I agree with the first reviewer.

Note: If any of the reviews include an additional attached file, it will not be attached to this message.   To access it, log in your Author Center and click on "Decision (View Letter)" -- this brings up this message, and at the bottom is the attached file. Click on that file name and you can see the attached comments from reviewers.